\documentclass[cameraready]{vgtc}

\ifpdf
  \pdfoutput=1\relax                   
  \usepackage{graphicx}                
  \usepackage{subcaption}
  \usepackage{enumitem}
\usepackage[table,xcdraw]{xcolor}
  \DeclareGraphicsExtensions{.pdf,.png,.jpg,.jpeg} 
\else
  \usepackage{graphicx}                
  \DeclareGraphicsExtensions{.eps}     
\fi%

\graphicspath{{figures/}{pictures/}{images/}{./}} 

\usepackage{microtype}                 
\PassOptionsToPackage{warn}{textcomp}  
\usepackage{textcomp}                  
\usepackage{mathptmx}                  
\usepackage{times}                     
\usepackage{cite}                      
\usepackage{tabu}                      
\usepackage{booktabs}                  
\usepackage{xcolor}

\onlineid{2294}

\vgtccategory{Research}

\vgtcinsertpkg

\title{SceneAR: Scene-based Micro Narratives for Sharing and Remixing in Augmented Reality}

\author{Paper ID:}

\author{Mengyu Chen\thanks{e-mail: mengyuchen@ucsb.edu}\\ %
        \scriptsize University of California Santa Barbara %
\and Andr\'es Monroy-Hern\'andez\thanks{e-mail: amh@snap.com}\\ %
    \scriptsize Snap Inc. %
\and Misha Sra\thanks{e-mail: sra@cs.ucsb.edu}\\ %
    \parbox{1.4in}{\scriptsize \centering University of California Santa Barbara}}

\abstract{Short-form digital storytelling has become a popular medium for millions of people to express themselves. Traditionally, this medium uses primarily 2D media such as text (e.g., memes), images (e.g., Instagram), GIFs (e.g., Giphy), and videos (e.g., TikTok, Snapchat). To expand the modalities from 2D to 3D media, we present SceneAR, a smartphone application for creating sequential scene-based micro narratives in augmented reality (AR). 
What sets SceneAR apart from prior work is its ability to share the scene-based stories as AR content. No longer limited to sharing images or videos, users can now experience  narratives in their own physical environments. Additionally, SceneAR affords users the ability to remix AR content, empowering them to collectively build on others' creations. We asked 18 people to use SceneAR in a three-day study, and based on user interviews, analyses of screen recordings, and the stories they created, we extracted three themes. From these themes and the study overall, we derived six strategies for designers interested in supporting short-form AR narratives. 
} 

\CCScatlist{
  \CCScatTwelve{Human-centered computing}{Visu\-al\-iza\-tion}{Visu\-al\-iza\-tion techniques}{Treemaps};
  \CCScatTwelve{Human-centered computing}{Visu\-al\-iza\-tion}{Visualization design and evaluation methods}{}
}

\begin{document}
\firstsection{Introduction}
\maketitle

Examples of early forms of storytelling with sequences of pictures can be found in Egyptian hieroglyphs, limestone carvings of Buddha, Greek friezes, Japanese scrolls, and illustrated Christian manuscripts, to name a few \cite{museum}. Over time, using picture panels and text for storytelling evolved into comics. More recently, with digital tools, everyday people can tell their stories through not only pictures and text but also sound and video (e.g., micro stories told with memes, GIFs, Snapchat lenses, TikTok videos etc.). SceneAR adds to this list of storytelling media by enabling the creation and consumption of micro narratives using smartphone-based augmented reality (AR). Unlike popular 2D media, creating micro narratives with AR and viewing them in the physical environment is an under-explored area and presents a rich set of research opportunities as well as unique challenges.

Although AR technologies have become more broadly available, tools for novices to quickly and easily create shareable narratives in AR are fairly limited. Mainstream development tools for creating AR experiences, such as Unity and A-Frame, often require extensive technical skills and programming knowledge~\cite{nebeling2018trouble}. The complexity of tackling AR-specific issues, such as surface detection, object registration, and tracking, can hold creators back despite help from ARCore \cite{arcore} and ARKit \cite{arkit} SDKs. Integrated solutions like Spark AR Studio by Facebook~\cite{spark} or Lens Studio by Snap Inc.~\cite{lens} are not as complex as Unity, but they are limited to desktop computers, require the use of additional 3D software tools (e.g., Maya and Blender) for creating 3D models, and expect the user to be familiar with programming. On the other hand, apps like Apple's Reality Composer~\cite{realitycomposer} and Adobe Aero~\cite{aero} have simplified the AR creation process for novices with the added ability to share an AR scene. But even these apps do not support creating or stitching together sequences of AR scenes nor do they allow remixing of AR content. 

In this work, we present SceneAR, a smartphone-based application that allows people to easily and quickly create and share micro narratives in AR. The stories are composed of sequences of AR scenes that contain 3D objects and text to form something like an AR comic strip. Although most mobile AR apps \textbf{compress 3D space into 2D} for sharing (i.e., turning an AR scene into an image or video), SceneAR enables viewing of the AR scene sequences by placing them in any physical environment (Figure~\ref{tab:comparison}). Users can freely explore the 3D models (characters, objects) in the scenes from any angle or distance by moving around and can interact with them using select, rotate, move, and zoom options. Furthermore, unlike existing AR apps, users can also modify and remix \cite{resnick2005some} AR scenes or entire narratives and share the updated versions. Remixing is defined as the \textit{``reworking and combination of existing creative artifacts, usually in the form of music, video, and other interactive media''}~\cite{dasgupta2016remixing}. Table \ref{tab:comparison} presents a summary of how SceneAR compares with existing AR apps and prior research that supports some form of story creation and consumption.
As a smartphone-based AR application, SceneAR not only enables ``immersive authoring''~\cite{lee2004immersive}, namely, the creation process that occurs in an immersive AR environment, but also in situ authoring~\cite{van2009situ}, namely, creation that happens in the same app used for consuming the content. We envision SceneAR could be used in areas such as storyboarding, stage design, social media, education, parent-children interactions, and creative arts. 
The main contributions of this work are:
\begin{itemize}[noitemsep]
    \item A smartphone-based app that enables non-expert users to easily create, share, consume, and remix scene-based micro narratives in AR.
    \item A new \textit{Micro AR} packaging format that integrates AR scene layout and content into a single shareable form. 
    \item Three themes derived from analysis of three-day user experience of SceneAR on creativity, spatial interaction, and remixing. 
    \item Six design strategies for designers and practitioners exploring AR for new forms of narratives. 
\end{itemize}

\begin{table*}[]
\resizebox{\textwidth}{!}{%
\begin{tabular}{@{}llllllll@{}}
\toprule
\textbf{} & \textbf{Creating Platform} & \textbf{Consuming Platform} & \textbf{\begin{tabular}[c]{@{}l@{}}In Situ \\ Authoring\end{tabular}} & \textbf{\begin{tabular}[c]{@{}l@{}}AR Sharing \\ Format\end{tabular}} & \textbf{\begin{tabular}[c]{@{}l@{}}Programming \\ Required\end{tabular}} & \textbf{\begin{tabular}[c]{@{}l@{}}AR Remixing \\ by Others\end{tabular}} & \textbf{\begin{tabular}[c]{@{}l@{}}Community \\ Repository\end{tabular}} \\ \midrule
Lens Studio + Snapchat & computer & Snapchat app & no & image or video & some & no & {\color[HTML]{3531FF}\textbf{yes, Snapchat lenses}} \\
SparkAR + Instagram & computer & Facebook AR Player & no & image or video & some & no & {\color[HTML]{3531FF}\textbf{yes, Instagram filters}} \\
Apple Reality Composer & {\color[HTML]{3531FF}\textbf{smartphone}} & \begin{tabular}[c]{@{}l@{}}Apple Reality \\ Composer app\end{tabular} & {\color[HTML]{3531FF}\textbf{yes}} & video & {\color[HTML]{3531FF}\textbf{no}} & no & no \\
Adobe Aero & {\color[HTML]{3531FF}\textbf{smartphone}} & Adobe Aero app & {\color[HTML]{3531FF}\textbf{yes}} & single AR scene & {\color[HTML]{3531FF}\textbf{no}} & no & no \\
Wonderscope & pre-authored & Wonderscope app & N/A & N/A & N/A & no & no \\
MagicBook (2001) & pre-authored & HMD + physical book & N/A & N/A & N/A & no & no \\
Magic Story Cube (2004) & pre-authored & mobile SceneAR app & N/A & N/A & N/A & no & no \\
ComposAR-Mobile (2009) & smartphone and desktop & smartphone AR player & no & N/A & yes & no & no \\
Educ-AR (2011) & desktop & desktop Educ-AR & no & N/A & {\color[HTML]{3531FF}\textbf{no}} & no & no \\
MARAT + ARCO (2013) & smartphone and desktop & MARAT app & {\color[HTML]{3531FF}\textbf{yes}} & no & {\color[HTML]{3531FF}\textbf{no}} & no & no \\
PintAR (2019) & \begin{tabular}[c]{@{}l@{}}specialized hardware + \\tablet \end{tabular} & PintAR Hololens app & {\color[HTML]{3531FF}\textbf{yes}} & video & {\color[HTML]{3531FF}\textbf{no}} & no & no \\
StoryMakAR (2020) & \begin{tabular}[c]{@{}l@{}}smartphone + browser + \\ specialized hardware\end{tabular} & StoryMakAR app & no & N/A & yes & no & no \\
SceneAR (ours) & {\color[HTML]{3531FF} \textbf{smartphone}} & SceneAR app & {\color[HTML]{3531FF}\textbf{yes}} & {\color[HTML]{3531FF}\textbf{multi-scene AR}} & {\color[HTML]{3531FF}\textbf{no}} & {\color[HTML]{3531FF}\textbf{yes}} & {\color[HTML]{3531FF}\textbf{\begin{tabular}[c]{@{}l@{}}yes, cloud-based \\ repository\end{tabular}}} \\ \bottomrule
\end{tabular}%
}
\caption{A summary of consumer applications and prior research focused on AR storytelling, both pre-authored and user created. The table shows how SceneAR differs from existing work in its ability to allow sharing sequences of AR scenes as micro narratives and its ability to remix shared AR stories. 
}
\label{tab:comparison}
\end{table*}

\section{Related Work}
SceneAR is inspired by prior work in short-form digital storytelling (e.g., memes, GIFs, TikTok videos), AR storytelling, and mobile AR authoring applications.

\subsection{Digital Storytelling}
Digital storytelling is a combination of narrative and technology used in many areas like entertainment, education, healthcare, and communication \cite{rossiter2010digital}. It emerged with the advent of accessible hardware and software, making it easy and fast for anyone to create and share their story. Digital storytelling is largely either interactive (e.g., video games, web documentaries \cite{ducasse2020, podara2021}, film \cite{bandersnatch}) or non-interactive (e.g., memes, gifs, comics, short videos) with some viral memes seeing recent sales as NFTs \cite{memenft}. 
Most non-interactive digital stories support meta-interactivity in the form of sharing, liking, and commenting. 
Comics have been a powerful medium of non-interactive storytelling for almost a century, exploring a variety of topics from daily life and political humor to tales of heroism and fantasy. SceneAR translates features of 20th century comics like panels, drawings, and dialog balloons~\cite{mccloud1993understanding} into AR as scenes, 3D models, and dialog bubbles. Each scene acts as an \textit{attention unit}, and narrative structure emerges when these scenes are placed in a deliberate sequence by the creator~\cite{cohn2013visual}. Taking inspiration from the French bandes dessinées to the Japanese manga and the American superhero comics, SceneAR proposes the idea of an AR comic experienced scene by scene in the user's physical environment.  

\subsection{AR Storytelling}
Prior work has explored AR for a variety of use cases like education \cite{billinghurst2012augmented}, creativity \cite{yilmaz2017using}, museum exhibits \cite{bimber2003virtual}, animation \cite{ye2020aranimator}, location-based experiences \cite{paay2008location,lee2012cityviewar}, tourism/heritage \cite{guimaraes2015augmented}, collaboration \cite{braun2003storytelling}, journalism \cite{pavlik2013emergence}, and audio-only experiences \cite{sra2013spellbound}. While there is an abundance of research on how to use AR, to our knowledge, a general purpose AR app for everyday use that allows users to create, share, and remix short-form narratives has not yet been explored.
Early works look at augmented print media that uses AR interfaces like MagicBook~\cite{billinghurst2001magicbook} and edutainment~\cite{grasset2008edutainment}. 
Magic Story Cube~\cite{zhou2004magic} introduced a tangible AR interface in the form of a folded paper cube where each unfolding produced different pieces of a story with non-animated visuals, voice, sound and music. 
ARFacade~\cite{dow2007ar} presented an interactive AR drama enabled by conversations with the characters. 
More recently, Kljun et al. \cite{kljun2019} explored the design space of AR-augmented comic books and examined how digital content can influence the behaviors of child readers. Wonderscope~\cite{wonderscope} introduced an interactive storybook app for kids with integrated voice recognition to encourage reading. 
StoryMakAR~\cite{glenn2020storymakar} is a recently introduced system that merges electro-mechanical devices and virtual characters that allow users to create stories with a browser-based block programming interface. MARAT \cite{ruminski2013creation} is an mobile authoring application that works with a back end tool that museums can use to design virtual exhibits.
Of these prior works, both StoryMakAR \cite{glenn2020storymakar} and MARAT \cite{ruminski2013creation} allow users to author AR content, though only the former allows the creation of a story using one of four virtual characters available in the app and custom hardware. Almost all other projects provide AR content already created by experts for consumption by non-experts. 
In contrast, SceneAR is an app that everyday users can use to create, view, share, and remix scene-based micro narratives with smartphone-based AR. 

\subsection{AR Authoring}
Prior work in AR authoring can be classified into two main types: low-level approaches that provide toolkits, libraries, or frameworks like ARToolkit, Studierstube, or Vuforia~\cite{kato2002artoolkit,schmalstieg2002studierstube,vuforia} and high-level approaches that are desktop-based authoring environments like Lens Studio, MARS, DART, or ComposAR~\cite{lens,macintyre2004dart,guven2003,unity,spark,seichter2008composar}. More recently, a third approach has become feasible with improvements in tracking and registration technologies, supporting AR content creation and consumption within the same AR system~\cite{aero,realitycomposer}. Feiner et al.~\cite{Feiner1997} developed one of the earliest mobile AR computing systems to allow spatially registered information for both indoor and outdoor locations to be accessed and managed by users through a combination of a see-through, head-worn display and a hand-held pen-based computer. Guven et al.~\cite{guven2006} proposed the \textit{Freeze-Frame} technique that enables the user to capture a snapshot of the environment and author AR content directly on top of it. Hengel et al.~\cite{van2009situ} presented a form of in situ authoring where 3D models generated from images could be inserted into the video stream. 
Langlotz et al.~\cite{langlotz2012sketching} presented an on-site authoring system that allows users to add 2D drawings and simple 3D objects in AR, though they do not offer any ability to scale, rotate, or move the virtual objects. 
Location-based game Tidy City ~\cite{wetzel2012tidy} allows end users to record and upload images and GPS locations to create AR scavenger hunt missions for other players. Zhu et al.~\cite{zhu2013authorable} proposed a bi-directional authoring tool that allows desktop users as well as maintenance technicians to author content on-site. Both of these approaches are similar in that they utilize pre-created AR environments that users can modify in a limited manner by taking new images or adding new text. None of these in situ authoring systems are real-time or immersive, namely, they do not allow for ``developing and experiencing the content concurrently throughout the creation process''~\cite{lee2004immersive}. In contrast, SceneAR is an immersive and in situ authoring tool that enables non-experts to create and experience AR scene-based narratives in a smartphone AR app. By providing access to thousands of 3D models, SceneAR makes authoring simpler and faster, much like Snapchat allows users to add AR lenses to their faces without having to create each lens from scratch.

\begin{figure}[!b]
    \centering
    \includegraphics[width=\columnwidth]{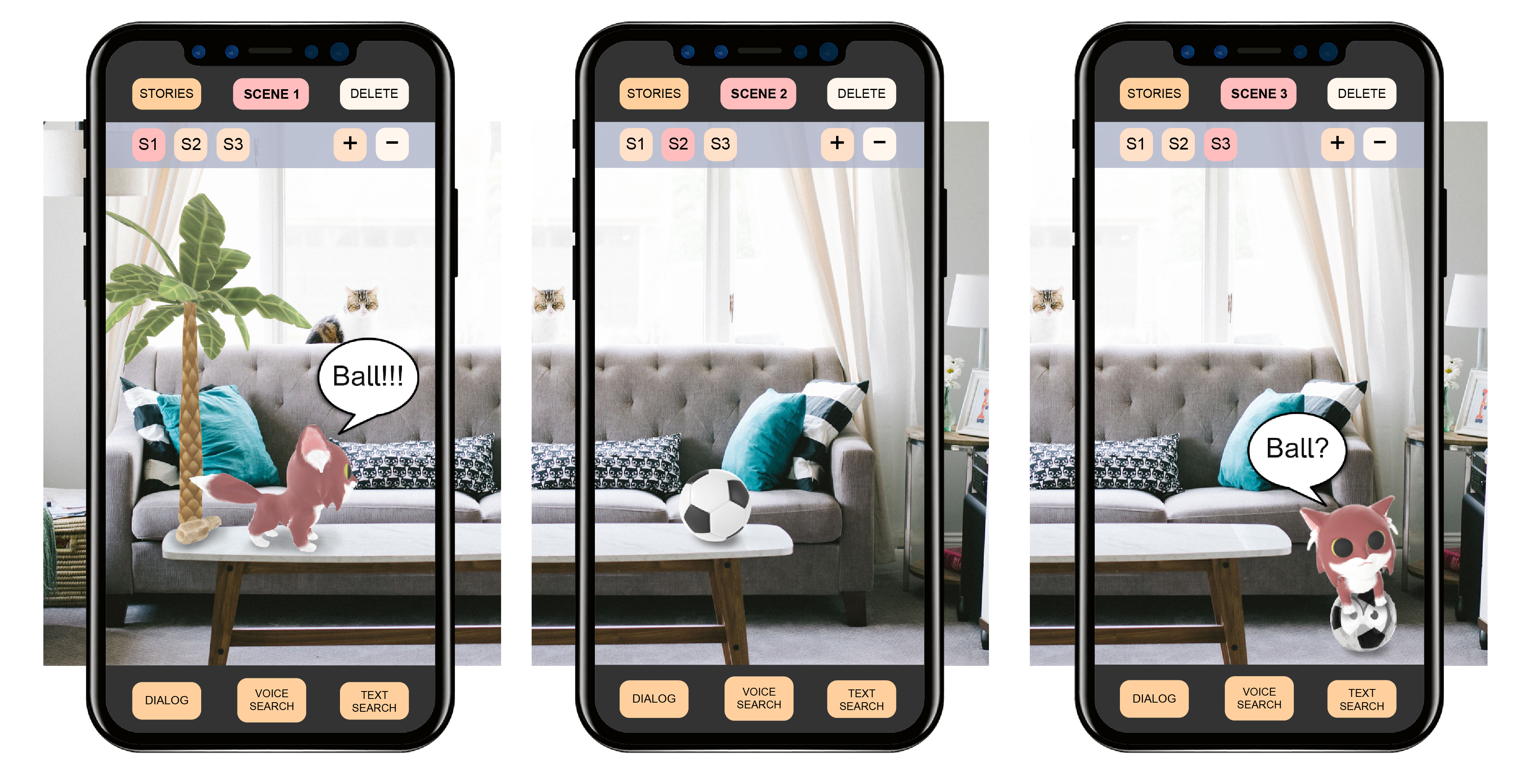}
    \caption{Concept design of a micro story told in three scenes where the user can view each scene on surfaces in their environment as a sequential 3D comic--like story made of three panels of 3D objects and dialog balloons.}
    \label{fig:my_label}
\end{figure}

\section{System Design}
\begin{figure*}[!t]
    \centering
    \includegraphics[width=0.9\textwidth]{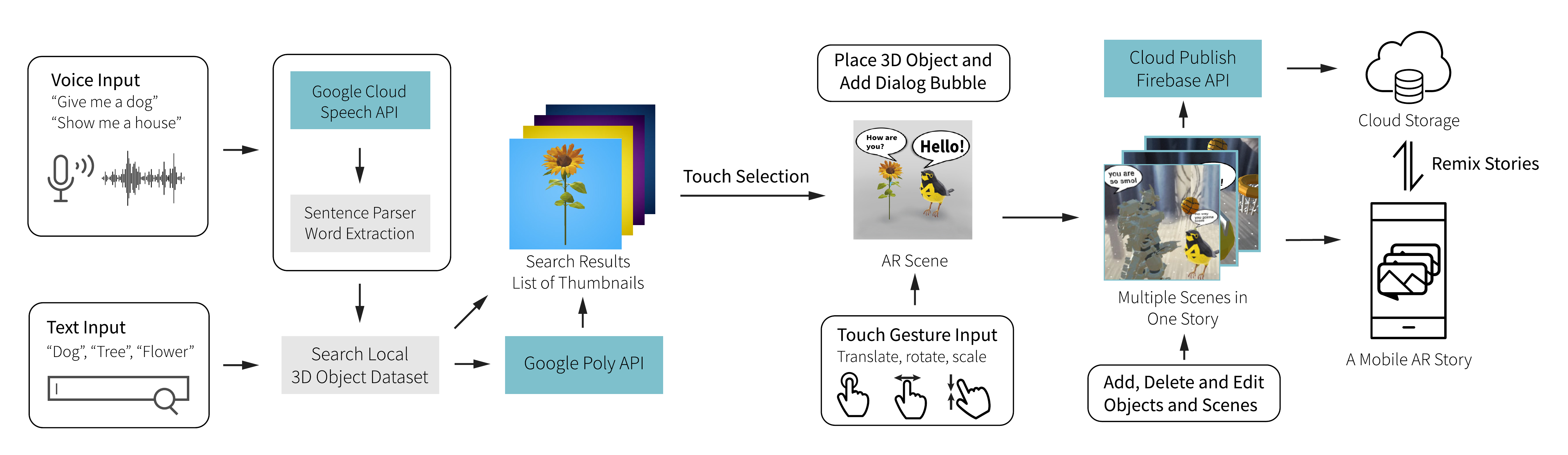}
    \caption{Overview of the SceneAR system (from left to right). Users search for 3D objects to add to an AR scene along with dialog bubbles. Multiple scenes create a sequential story. Published stories are downloaded and viewed in AR. Anyone can remix any story by adding or removing 3D objects, scenes, and dialog bubbles.}
    \label{fig:system}
\end{figure*}

We built SceneAR in Unity as a mobile Android application using Google's ARCore API~\cite{arcore}. ARCore allows SceneAR to track the position of the phone in physical space and build an understanding of the user's surroundings by detecting planar surfaces in the environment. 
Figure~\ref{fig:system} shows an overview of the SceneAR system.

\subsection{Narrative Components}
\paragraph{Scenes}
A micro narrative in SceneAR is defined as a collection of AR scenes, each composed of 3D models that the creator or viewer can place on surfaces in their physical environment. Each scene is modified individually, and users can select, add, or delete scenes as needed with no limitation on the number of scenes in a narrative. Inspired by comic strips, our scene-based design enables a sequential narrative in AR. 

\paragraph{Interaction}
Interaction with SceneAR has two elements: touch and speech. Touch-based interaction is used for AR object manipulation and for interaction with the app UI. SceneAR recognizes the following touch gestures: 1) tap for object placement and selection, 2) drag for object translation, 3) twist for object rotation, 4) pinch for object scaling, and 5) two-finger drag for object elevation. Speech-based input is used for 3D object search and for creating dialogues as an alternative to text-based search and dialog creation. 

\begin{figure}[!b]
    \centering
    \includegraphics[width=0.9\columnwidth]{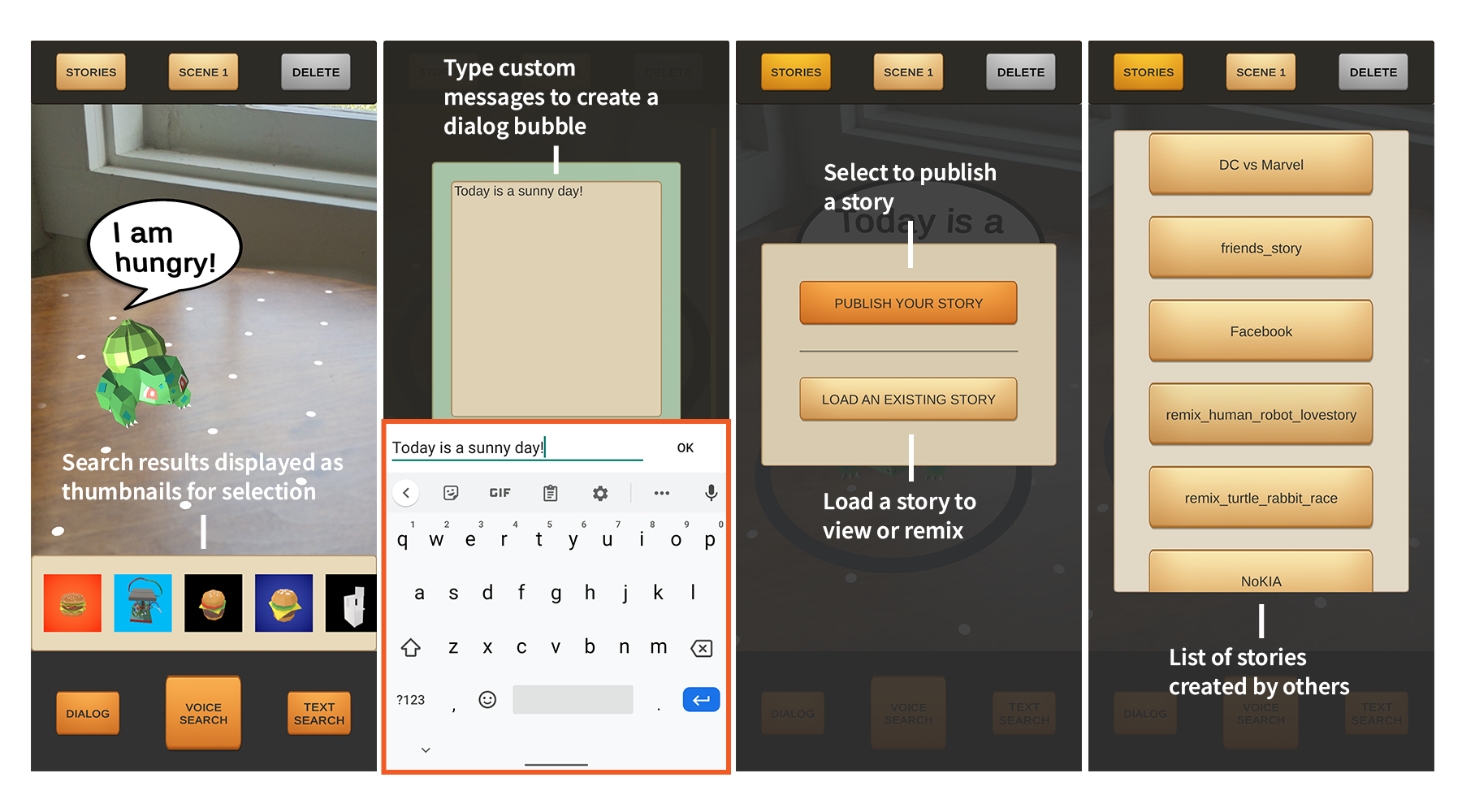}
    \caption{Figure showing the main SceneAR scene-creation elements. From left to right: 3D object search results displayed as a scrollable list of preview icons at the bottom. A custom dialog can be added via typing or speech-to-text input. Stories can be published or loaded for viewing/remixing. Browsable list of all stories is available for viewing and/or remixing.}
    \label{fig:interface}
\end{figure}

\paragraph{Authoring and Presentation}
Users have access to a large set of 3D models in the SceneAR app through the Google Poly 3D object library~\cite{poly}. These objects can be accessed via voice or text search (Figure~\ref{fig:system}). We included speech input because users are already familiar with using Siri or the Google Assistant to interact with their smartphones. Google Cloud Speech API parses speech data to text, which is then used to return a set of 3D models from Google Poly (Figure~\ref{fig:interface}). 
The app also includes a smaller set of animated humanoid models to allow creation of scenes if the user is temporarily offline. Figure~\ref{fig:interface} shows the main steps for creating an AR scene. 
There is a separate module to add dialog balloons to any object in a scene. The user can choose from pre-existing dialogue or create their own using speech or text. The dialog balloons can be removed or edited for remixing.

\subsection{Application Elements}

\paragraph{Publishing, Viewing, and Remixing a Story}
Before publishing (Figure~\ref{fig:system}), users are asked for a title and a description, including any physical requirements (e.g., best placed on the bath tub) that potential story viewers, who can freely place the scenes anywhere in their environment, should know. Once a scene is placed, the viewer can edit it directly by adding/deleting 3D objects, dialog balloons, or entire scenes to modify or expand upon the narrative. This modified or remixed story can be published with a new title and description. 

\paragraph{Micro Narrative Packaging}
SceneAR allows publishing and sharing 3D comic-style stories. To enable this, we created \textit{Micro AR}, a new container format for packaging scene-based AR stories. Inspired by the open-source OpenDocument Format (ODF) that packages a text document from different components like metadata, content, and formatting~\cite{openDocument2015}, the \textit{Micro AR} format is designed to package three main components of the AR story: 1) metadata, 2) story content, and 3) spatial layout.

\noindent \textit{Metadata}
The Micro AR format contains app-generated and user-created metadata about the AR story. It includes fields such as creator, title, description/viewing guidance, original creator, creation timestamp, and viewing statistics. The metadata defines and differentiates each AR comic as a unique document. Metadata allows the user to browse through a large library of stories before deciding which one to load and view.
\\
\\
\noindent \textit{Content}
Content contains both local and remote data related to each story. Local data are mainly dialog texts created for each scene in a story. These are small in size and packaged for sharing directly. Remote data are content-specific metadata that include 3D asset names and unique keys. Because these assets are large in size, the app uses keys to retrieve the actual 3D models from cloud storage when needed. The assets are downloaded in the background as the user scans their environment to make the story appear immediately upon plane detection and selection for an improved user experience.
\\
\\
\noindent \textit{Spatial Layout}
All content in an AR story is placed on a physical surface for creating and for viewing. Thus, the Micro AR package needs to track the user-defined spatial layouts in each scene of a story. Positions, rotations, scales, and groupings of 3D objects relative to each other in each scene are recorded and packaged so that each reconstruction and viewing of an AR story has the same spatial layout as the original creator intended. 

\section{Three-Day User Study}\label{sec:eval}

We evaluated SceneAR in a three-day field study with 18 remotely located participants. Before the study, we ran a pilot study with two remote participants as a dry run of the study and to get early feedback on the app's functionality for identifying and fixing technical issues.

\subsection{Participants}
Eighteen participants were recruited through word of mouth and mailing lists of various academic departments around the United States. Because we developed the SceneAR app using Google's ARCore SDK, we looked for participants who owned phones officially supported by ARCore. Participants were located in six different states of the United States, Brazil, India, and China. Participants included graduate and undergraduate students, engineers, software developers, and a chemist (12 male, 6 female), aged between 20 and 44. 

Participants rated, on average, 4.1 on a 7-point Likert scale (1 = never before, 7 = a great deal) their familiarity with AR. Pok\'{e}mon Go was the most common AR experience cited. Participants reported spending an average of 6.7 hours per week using personal devices like phones, tablets, or computers and 1.7 hours a week playing games on the phone or tablet.
Participants also reported posting visual content frequently on WeChat (2), Snapchat (7), and Instagram (9), including memes (6), GIFs (3), comic strips (1), and YouTube videos (2). Two participants reported rarely using social media.

\subsection{Study Procedure}

\subsubsection{Pre-study} Because ARCore is only supported on recent devices, finding participants with official ARCore-approved devices was challenging. Therefore, potential participants were asked to install Measure (an AR app) and AZ (a screen-recording app) from the Google Play Store, record a session running Measure for 3 to 5 minutes, and report on device overheating, excessive battery drain, and any Measure app crashes. Our goal was to ensure ARCore could run on devices at least well enough for participants to successfully complete the SceneAR study tasks, even if they were not officially supported. Despite this test, some participants had severe issues with ARCore surface detection (see Results~\ref{questionnaire}).

\subsubsection{Onboarding} Each participant was onboarded in an online meeting using Zoom video conferencing software. This meeting lasted 40 to 50 minutes. At the beginning of the video call, we emailed the participants a consent form (study protocol approved by our office of research), a participant ID, a pre-study questionnaire, the app installer (APK), instructions on how to install the app, a document with information about the study tasks and how to use the app, and a drive link to upload their screen recordings. During the video call, we walked participants through all these items. Participants also screen recorded themselves creating a story and remixing a story during the Zoom meeting.
At the end of the Zoom session, participants were asked to use SceneAR as they pleased with a study task requirement to create a total of nine stories over the three days. We closed by scheduling a follow-up Zoom meeting three days later for filling out a post-study questionnaire, conducting a short interview, and showing participants how to remove the app from their devices.

\subsection{Data Collection}
Data were collected through screen recordings of app usage sessions, published stories, pre- and post-study questionnaires, and Zoom recordings of interviews. Because of the open and remote nature of the study, we asked participants to screen record all their app usage sessions and upload them to a secure folder at the end of the study.
After the three-day study, participants were interviewed using a semi-structured interview schedule, which lasted an average of 20 minutes. Before that, participants were asked to fill out a post-study questionnaire with two 7-point Likert scale (1 = strongly disagree, 7 = strongly agree) questions, along with questions about what they liked/disliked about SceneAR, the types of stories they enjoyed creating, barriers to using the app, and other open-ended feedback.

\section{Data Analysis}\label{sec:results}
We wanted to identify and quantify SceneAR usage, so the data were thematically coded by two researchers independently to determine characteristics and functionalities.
We employed an inductive thematic analysis approach to the data, as described by Braun and Clarke \cite{braun2006using}. Participant interviews were transcribed from the Zoom recordings using an online service (sonix.ai), and the text transcripts were exported for qualitative analysis. Screen recordings were matched with interview transcripts using the participant IDs. Two researchers independently reviewed these transcripts and recordings. 
To make meaning from the screen recordings and the interviews, each researcher created their own codes. Following that, the researchers met over Zoom to discuss and refine their codes, which resulted in an agreement on 12 codes. These codes were further analyzed and referenced with the screen recordings and feedback in the questionnaires and then reviewed again by both researchers in another meeting. This approach identified three overarching themes, as discussed below. 

\begin{figure}[!t]
    \centering
    \includegraphics[width=0.9\columnwidth]{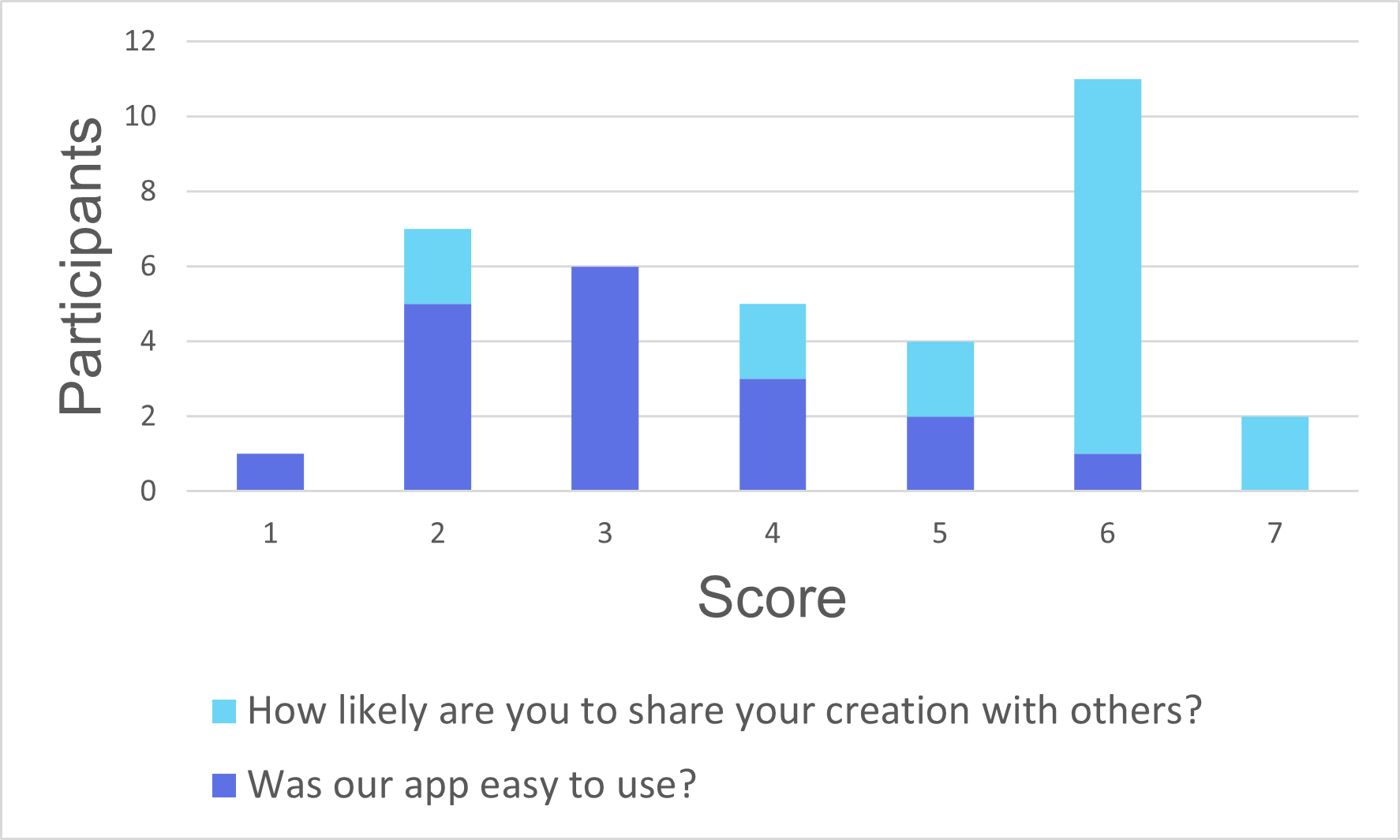}
    \caption{Participant (N = 18) responses to two questions in the post-study questionnaire asking 1) how likely are they to share a created micro story (1 = not at all, 7 = extremely likely), and 2) was the app easy to use (1 = very easy, 7 = very hard).}
    \label{fig:questionnaire}
\end{figure}

\section{Results}
In this section, we present the responses to the post-study questionnaire and describe the three themes we derived from our analysis of the data: Diversity in Creativity; Spatial Creation and Viewing; and Sharing, Remixing,
and Collaborating.

\subsection{Questionnaire Responses} \label{questionnaire}
Likert responses are illustrated in Figure~\ref{fig:questionnaire}. Most of the participants found our app easy to use (1 = very easy, 7 = very hard), with a positive response median (\textit{M}) of 3 and a median absolute deviation (\textit{MAD}) of 1. Participants also agreed that they were likely to share their creations with others (1 = not at all, 7 = super excited to share) with \textit{M} = 6, \textit{MAD} = 0. To understand the reasons for these scores in the interview data, we found that two participants who rated our app as being not very easy to use had frequent ARCore-related surface detection issues on their devices. One user had an old phone model that did not provide a stable frame rate, and the other user was detecting surfaces under extremely poor lighting. On likelihood to share the creations with others, two participants rated it low, which matched their prior report on rarely using social media, with one saying, \textit{``[Do] not post very often and pretty much never publicly share things like gifs or memes."}

\begin{table}[b!]
     \centering
     \begin{subfigure}[b]{0.50\textwidth}
         \centering
         \begin{tabular}{@{}lll@{}}
\toprule
\textbf{App Usage Data Summary}                         &            \\ \midrule
Published stories   & 194                 &\\
Remixes             & 48                  &            \\
Self remixes        & 11                  &            \\
App usage time per user (average)         & 67 min          &       \\
Sessions per user (average)       & 11        &   \\
Session duration (average)        & 20 min      &           \\
Unique 3D objects used            & 325         &            \\
Total 3D objects               & 1,204                &            \\ \bottomrule
    \end{tabular}

\end{subfigure}
     \caption{Summary of SceneAR usage during the study.}
    \label{tab:appdata}
\end{table}

\subsection{SceneAR Usage Summary}
A summary of how participants used the app is presented in Table~\ref{tab:appdata}. Participants used SceneAR in bursts throughout the study, averaging about 20 minutes of app usage per participant per day. The minimum time spent in one session was about 10 minutes, whereas the maximum time was about 34 minutes. The minimum total app usage time over the three days was about 30 minutes, whereas the maximum was over 3 hours.

Across all 18 participants, a total of 194 stories were created, of which 25\% (48) were remixes. Of the original stories, 26\% had one scene, 32\% had two, and 42\% had three or more scenes. \textit{``Mary had a little lamb''} by P17 had the most scenes of any story at 10, whereas \textit{``Lonely\_Cat''} by P5 was the second longest at 8 scenes.
The longest remixed story had six scenes, titled \textit{``remix\_don\_quixote''} (Figure~\ref{fig:don}), originally created by P5 with five scenes and remixed by P14 with a new scene. Of the remixed stories, 22\% were remixes of stories created by the participant themselves, whereas the rest were remixes of other people's stories. The most used search keyword was ``person'' for a total of 142 times not including other related searches for ``old man, girl, black woman, athlete, drummer, artist, Trump, etc.'' and fictional character searches like ``Luke Skywalker, Iron Man, Mario, Captain Marvel, Darth Vader, Gandalf, etc.'' The top three story themes were COVID-19, music, and space.

\begin{figure}[!t]
\centering
\begin{subfigure}[b]{.48\columnwidth}
\centering
\includegraphics[width=1\columnwidth]{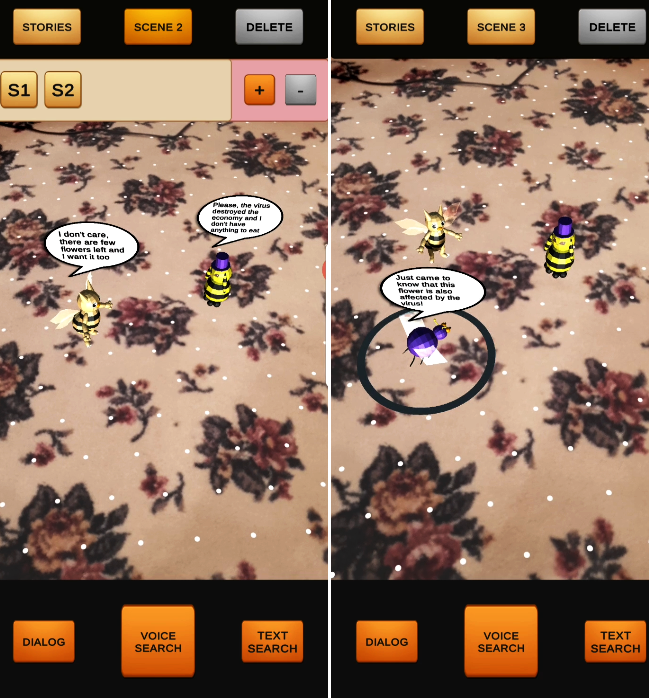}
\caption{\label{fig:bees}}
\end{subfigure}\hspace{1mm}
\begin{subfigure}[b]{0.48\columnwidth}
\centering
\includegraphics[width=1\textwidth]{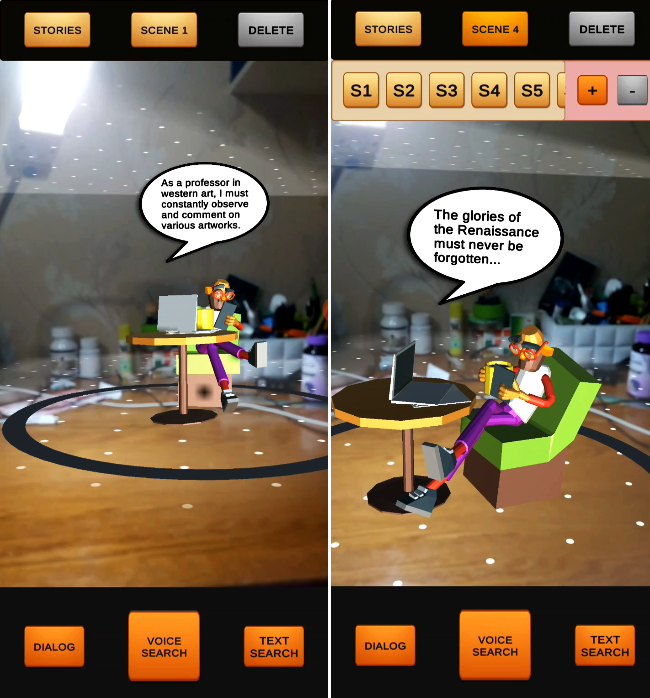}
\caption{\label{fig:prof}}
\end{subfigure}
\caption{a) Scene one shows two virtual bees arguing about who gets the flowers on the physical rug because of the shortages created by the COVID-19 pandemic. A third bee joins them in Scene 2 to end their argument by saying the flowers are infected by the virus. b) Scene one shows an art history professor lecturing. The two scene screenshots show how the creator framed the professor, by scaling the model instead of moving the camera, inspired by their background in film making.}
\label{fig:bees-prof}
\end{figure}

\subsection{Theme 1: Diversity in Creativity}\label{sec:DC}
This theme describes how participants were able to create the stories from various sources by using different narrative techniques. It is divided into four categories: Moments from Life, Media Inspiration, Improvisation on Visual Search, and Technique Transfer. We found that participants were eager to discuss the types of stories they created and why, requiring little to no prompting to describe their experience. Several participants expressed a desire to continue using the app after the study. 
\\
\\
\noindent\textit{Moments from Life} \\
A large number of stories created by participants depict moments from life, also commonly seen in digital storytelling media like memes, GIFs, and short-form videos. They cover topics such as home, relationships, recreation, politics, games, nature, etc. Among these daily life--themed stories, a variety of ongoing social events including the COVID-19 pandemic, space travel, and government policies (the international student ban, shelter-at-home orders, etc.) can be observed. COVID-19 stories specifically focused on the economic crisis, social distancing, mask wearing, Christmas without Santa, travel bans, and loneliness. For example, P8 created a digital-physical story in three scenes called \textit{``bees in covid''} that showed virtual bees standing on a physical floral print rug and having an argument about who should get the flowers (Figure~\ref{fig:bees}) in this time of crisis and shortage.
\\
\\
\noindent\textit{Media Inspiration} \\ 
Participants narrated how much they enjoyed re-creating content from other media like movies and storybooks. The \textit{tortoise and hare} story created by P4 was remixed by two others with new scenes and dialogues. P6 said, \textit{``We can recreate remixes of different, like, stories that we've read before like Cinderella comes and picks up her shoe. Having a twist, maybe something is not similar, at the same time it's a visual depiction of that [something familiar]''}. P6 was interested in creating old stories that teach philosophy and morality. P7 created a space battle based on the movie franchise \textit{Star Wars}, showing virtual ships and two 3D-printed ships, namely the Millennium Falcon and a TIE Fighter. The \textit{Don Quixote} story was created and remixed in a playful way by several participants (Figure \ref{fig:don}).
\\
\\
\\
\noindent\textit{Improvisation on Visual Search} \\
Participants improvised based on the app's responses to their search results supported by access to thousands of 3D models, though sometimes even that large set did not seem enough.
We noticed that, during an object search, the selected 3D object does not always match perfectly with the intended keyword input. Some cases were due to the absence of a 3D model that matches the participant's expectations to use as a story component, which is a limitation of keyword-based search methods. However, in some cases, the participants were inspired by the visual results of a search and decided to dedicate the story plot to the found 3D object. This can be particularly observed in the story creation process by P13 where the story was originally intended to be a basketball game between Pikachu and Godzilla. P13 initially fetched a 3D model of a regular basketball but later found a spherical-shaped ball-like Pikachu while trying to fetch a regular Pikachu 3D model. This made P13 change the story plot into an ending where Pikachu becomes the ball in the basketball game, thus deleting the original basketball. Similar behaviors can also be observed in P16's creation process where one of the main characters of the story switched from a giraffe to a fox during the visual search of 3D objects.
\\
\\
\noindent\textit{Technique Transfer} \\
Participants used their prior experience with other forms of media to help guide their 3D creative experience.
A story by P5 composed of seven scenes titled \textit{``On\_Art\_History''} is a monologue by a professor of Western Art speaking to his audience (presumably lecturing in class) (Figure~\ref{fig:prof}). We noticed each scene was explored in a manner similar to framing shots in a film, something P5 expressed during the interview. About framing scenes, P5 said, \textit{``Even though I did not use any real world objects, I was still able to enjoy the functionality of using different angles and sometimes making it big, sometimes make it small, sometimes make it distance, sometimes make it close.''} P4 said, \textit{``I'm into short films. This app, there are more 3D models and objects. I felt like it\'s more useful for future filmmakers and all to set their scene in this AR thing so they can show it to their crew and team member  to like better understand director's vision.''} P1 enjoyed placing objects in scenes and said, \textit{``You are actually constantly a theater designer or stage designer. That's already like a whole another set of entertainment for me.''} P4, P12, and P7 used onomatopoeic words commonly seen in comics (e.g., 'Raawr', 'VROOM', 'Zzzz', 'SCREECH', 'pew pew') and emoticons seen in text messaging in their dialog balloons to express emotions. 
\\
\\

\begin{figure}[!t]
    \centering
    \includegraphics[width=0.9\columnwidth]{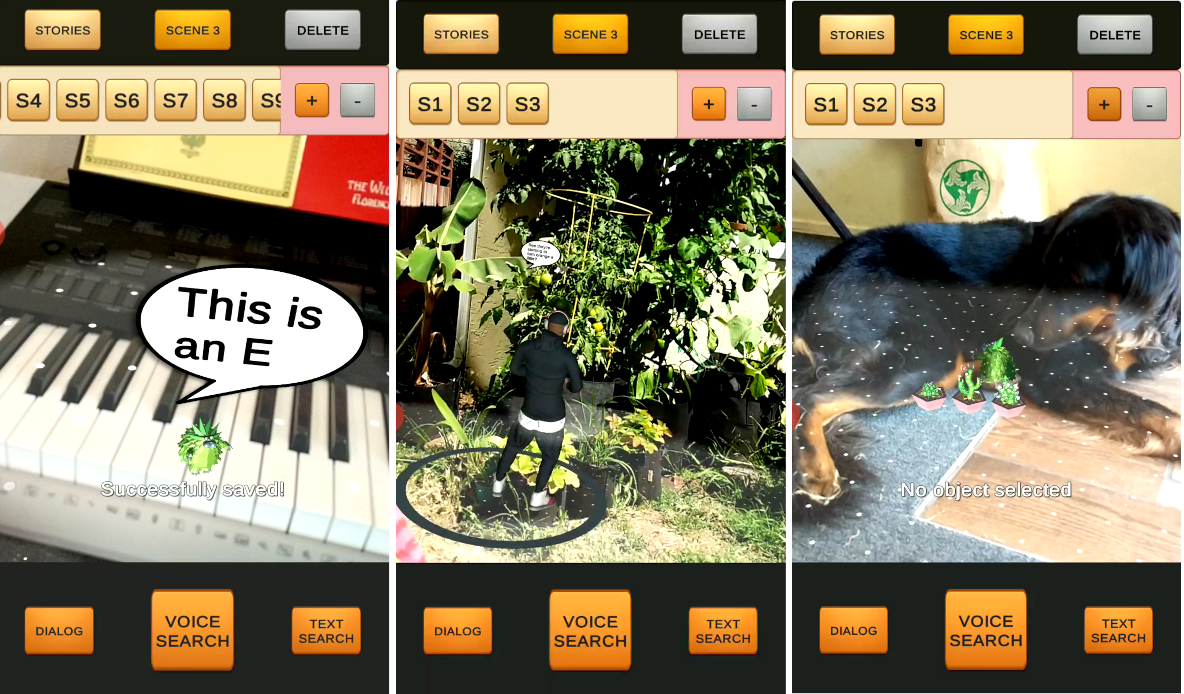}
    \caption{Three micro stories created by P18 that connect virtual objects with the real world. Left: A 10-scene narrative that teaches how to play ''Mary had a little lamb'' on the piano. Middle: A story set outdoors where the character explains how to recognize different tomato plants. Right: The dog being mildly annoyed by virtual bugs. }
    \label{fig:ptd}
\end{figure}
\subsection{Theme 2: Spatial Creation and Viewing}\label{sec:SC} 
\begin{figure}[!t]
\centering
\begin{subfigure}[b]{.47\columnwidth}
\centering
\includegraphics[width=1\columnwidth]{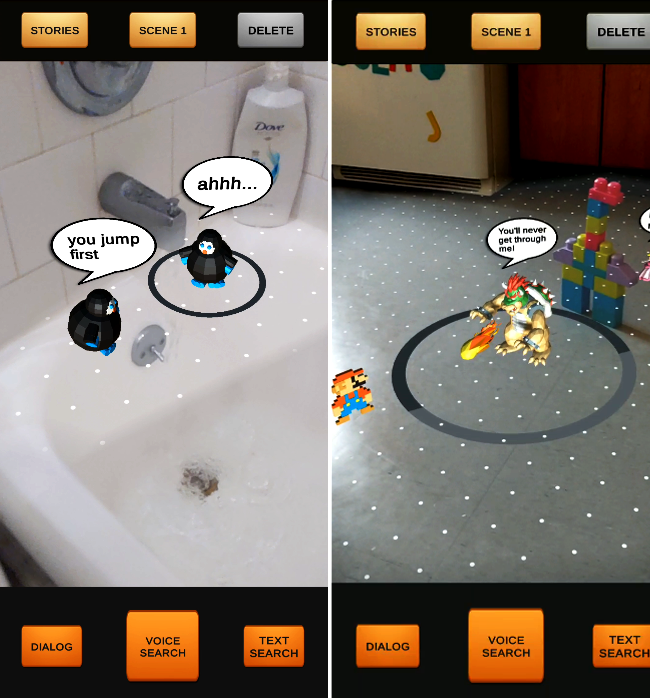}
\caption{\label{fig:tub}}
\end{subfigure} \hspace{1mm}
\begin{subfigure}[b]{0.48\columnwidth}
\centering
\includegraphics[width=1\textwidth]{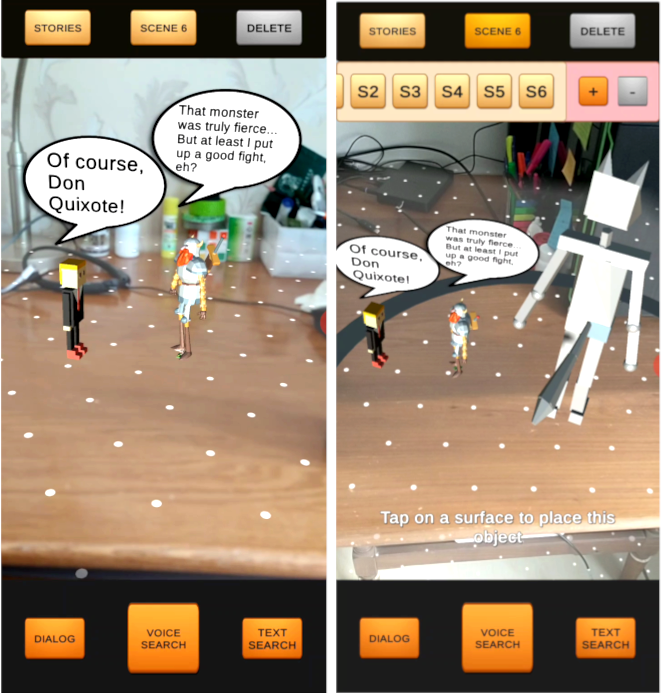}
\caption{\label{fig:don}}
\end{subfigure}
\caption{a) Left: two penguins debating who jumps first while standing on the edge of a bath tub with the water running. Right: a re-creation of the Super Mario video game with Princess Peach hidden behind the physical toy castle and virtual Bowser defending it. b) This set of two screenshots shows a scene from the original story \textit{''Don\_Quixote''} by P5 on the left and the same scene remixed by P14 with the addition of a robot cat on the right.}
\label{fig:mario-veg}
\end{figure}

This theme focuses on creating and consuming stories in a 3D spatial environment. We divided it into four categories:
In Situ Authoring for Creativity, Spatially Stimulated Imagination, Story Setting, and Physical and Virtual Object Interplay. We found that our users specifically enjoyed the simplicity and speed of the scene creation process and did not remark upon or ask for inclusion of other features (sound, animation control, events, etc.) as necessary elements for telling their Micro AR stories.
\\
\\
\noindent\textit{In Situ 3D Authoring for Creativity}\\
Participants expressed surprise and excitement about being able to create 3D content using their mobile devices. They reported that SceneAR enabled free-form creativity in AR, with 3D adding another dimension and enabling new possibilities that they had not yet seen in any other app. P5 said, \textit{``It is just so cool, like you create some 3D scenes with dialog with different scenes, on a mobile device, in 20 minutes.}'' Some found 3D to be the next obvious step saying (P5),\textit{``You create from drawings to moving frames, movies, then there are just more things to do. Right? I think there will be more storytelling ways, possibilities in this new platform.''} Participants found the ability to express themselves in 3D liberating. P6 said, \textit{``Having it in 3D just expands the scope. In 2D its just an image and it is quite constrained but in 3D like we can have like different scenes so it just expands the scope.''} P12 remarked upon how stories in 3D \textit{``become part of your world''} while P15 said the 3D aspect was \textit{``definitely part of the appeal. There isn’t really anything that is creating in the app in 3D and viewing in the app in 3D.''} On 3D authoring enabling greater creativity, P4 said, \textit{``We are going from 2D world to a 3D world and creating stuff. In 3D world a lot of other stuff can be include which we can’t put on paper. It will help people be more creative and like convey the message that they want a little more clearly to others.''} P17 said 2D has barriers but with SceneAR everything became much more limitless, though P3 worried that there may come a time when there are \textit{``too many layers on top of what you’re seeing and you are losing sight of what’s really there.''}
\\
\\
\noindent\textit{Spatially Stimulated Imagination}\\
The ability to create virtual content with changing perspectives and realistic life-size scales seemed to have presented the participants with a new mental model for thinking about stories. In AR, stories and storytellers inhabit the same space, which is a blend of physical and virtual elements. Some found this new relationship helped them think of new ways to tell their story. P7 said, \textit{``I found myself thinking of special stories content that I could create that I wouldn’t otherwise be able to create using other platforms.''} P8 expressed a similar idea about the scale of objects: \textit{``This is like an environment where you can make anything kind of happen, like you can make robots talk or like flies be bigger than dinosaurs.''} Participants found the ability to change perspectives a new way of experiencing a story. P3 said, \textit{``[O]nce I realize I can move and see it from different angles and its like that, that helped coz you can be inside it from different angles where you sorta like can’t with other ways like drawing or whatever, you can’t be like inside it!''}
\\
\\
\noindent\textit{Story Setting}\\
We noticed participants created their stories in different settings around their homes---from bathrooms, kitchens, and hallways to bedrooms and front yards. 
P15 created a story called \textit{``penguins go swimming''} where virtual penguins debate who will jump first while standing on the edge of a physical bath tub (Figure~\ref{fig:tub} left). P15 created another story on the kitchen counter introducing a virtual cookie to a physical one. Understandably, these stories require the viewer to have similar physical conditions to experience the story as designed.
When asked about blending the real and the virtual, P15 said, \textit{``[It] is a really fun and exciting new mechanism for people to share their creativity.''} P14 echoed that sentiment, saying, \textit{``I think it's the opportunity to mix real world with fiction AR.''} P18 was the only participant who made stories outdoors, one story showing their vegetable garden, another one teaching the viewer how to play a song on the piano, and yet another one attaching virtual bugs to their dog (Figure~\ref{fig:ptd}). When asked, P18 remarked, \textit{``Every new tool is a new way to think about things and do things in different ways.''}
\\
\\

\noindent\textit{Physical and Virtual Object Interplay}\\
In addition to setting a story in a physical place (e.g., kitchen counter), we noted that participants created stories with physical objects.
P7 created a fictional story based on the popular video game franchise Super Mario using the game characters and a physical toy castle (Figure~\ref{fig:tub}, right) as the main elements. P6 expressed a desire for more interaction with physical objects saying, \textit{``If I could use dialogues on physical objects, even though there was no object detection, I could just put a dialog.''}
P7 said their favorite part was \textit{``making the digital and physical worlds act with each other.''} These participants demonstrated a different type of relationship between the physical and the virtual, one where the setting has no relationship to the story but the story itself integrates physical and virtual elements, for example, a toy castle with the virtual Mario (Figure~\ref{fig:tub}, right). P7 created two stories (\textit{``Mario''} and \textit{``Star Wars''}) that incorporated physical objects like toys and 3D-printed space ships, but not the physical location, into the stories.
\\
\\
\subsection{Theme 3: Sharing, Remixing, and Collaborating}
This theme relates to participants being able to share and remix stories made by others in AR, unlike anything they reported having seen or tried before. Our new \textit{Micro AR} story packaging method enables easy sharing and remixing of the user-created AR scenes. The four categories are: Stories for Sharing, Remixing, Inspiration from Community, and Communication and Collaboration.  
\\
\\
\noindent\textit{Stories for Sharing}\\
A fundamental aspect of storytelling is the audience. All participants in our study enjoyed creating and sharing stories as reported on the post-study questionnaire. P7 likened people using SceneAR to those creating short-form video stories on TikTok~\footnote{\url{https://www.tiktok.com/en/}}, saying 
\textit{``[P]eople that are good at making shorter funny content that would have a blast with adding these [3D] new elements into the scene.''} The desire to include friends in stories both directly as avatars and indirectly as people to share stories with was expressed by several participants. Some participants added characters that resembled their friends, and some expressed a desire for the app release to include a feature that would allow them to import photos of their friends into the 3D scenes. 
\\
\\
\noindent\textit{Remixing}\\
The idea of collaborative remixing~\cite{dybwad2005approaching} is evident in participants creating stories specfically for others to remix. For example, P8 says, \textit{``I made it [a story about wormholes] so that other people would have like good remix ideas for it.''} All participants remarked upon the novelty of remixing AR scenes and how much they enjoyed it. P11 was the only participant who did not create any remixes, being one of two people (other being P3) who also mentioned not posting on social media.
Nevertheless, P11 created the most stories (13) of anyone in the study. We would like to believe that they found their medium of creativity in SceneAR. P1 also did not create any remixes saying, \textit{``I never felt like I have a motivation to trying to do it. Like storyline is so complete a lot of time that it's better to just start over sometimes.''}
Six participants explicitly mentioned enjoying seeing remixes of their own stories or creating their own remixes, which we call self-remixes. P2's remixed story presented a continuation of one idea from the first story to the fourth one with each new story adding new scenes, characters, and objects (Figure~\ref{fig:don})
They said, \textit{``[Remixing] is really cool, like seeing other people part of your story.''} Of those who enjoyed remixing, P8 said, \textit{``Remixing stories was like a lot of fun, yeah, that was definitely very interesting.''}
\\
\\
\noindent\textit{Inspiration from Community}\\
During the interview, participants highlighted that the ability to remix made creating stories much easier. One participant contrasted the mindless ease of taking and sharing a photo on Instagram with putting a lot more care and purpose into the design of their scenes. P5 said, \textit{``Most of the time I would start from seeing somebody else's work. It would give me inspiration. If it's really good I would just expand on that. I see it as a source of motivation, giving you ideas.''} P10 said, \textit{``We can follow the idea from other people and then build on top from it.''} Three participants said coming up with creative or clever ideas was the hardest part of using the app. P4 said, \textit{``For me, was the creativity to create the stories.''} P14 also said the creativity was the hardest part for them and being able to remix was helpful: \textit{``You can pick idea of another person and improve this, so I think is very nice.''} P1 thought the app allowed open-ended creativity like Lego blocks but, unlike Legos, where you are constrained by the pieces you have, SceneAR gives you \textit{``unlimited 3D objects, unlimited resources to create your own work.''} P17 wanted the app to give them suggestions for objects \textit{``based on the physical space recognition''} to make the experience more intuitive, such as creating stories related to living rooms if they are in the living room. 
\\
\\
\noindent\textit{Communication and Collaboration}\\
We noted that remixing afforded a novel form of back-and-forth communication between participants where one person expressed their thoughts and emotions through a story and another person remixed it only to be remixed again by the first person. P2 remarked, \textit{``We don't want to build everything from scratch. I need some existing background that I can put my people and you know start like conversation.''} P12 found remixing a good way to \textit{``share imagination and creativeness with each other.''} P12 said that seeing other people modify your stories gives you an understanding of how they have interpreted your work. During the interviews, participants expressed a desire to have a social network and the ability to create these back-and-forth narrative-based dialogues with friends. P6 said, \textit{``if I could have friends on this app and I can share the story and they could render it in their surroundings, like they could load it in their surroundings, then I can like ask them to edit, so it would be like a collaborative creation.''}
\\
\\

\section{Discussion: Strategies for Supporting AR Storytelling}
Here we articulate six design strategies that we derived from our data analysis and participant feedback. These strategies can serve as a guide for the development and design of future AR short-form storytelling systems that allow users to create, share, and remix 3D scene-based narratives on their mobile devices.

\subsection{Consider Spatial Dependencies}
One of the challenges with sharing AR stories is that the creator's physical environment is likely to differ from the consumer's. For example, an AR story can have contextual dependencies, such as the size of the detected plane, which might impact the position and rotation of AR objects. This could happen if, for example, the creator uses a large dining table to create the story and the viewer tries to render it on a small coffee table. 
Similarly, a creator might compose a story on multiple detected planes. For example, a creator might create one scene on the floor and a subsequent one on a table. If the person viewing the story does not have an environment with two planes at different relative heights, some virtual objects might appear to be floating.
These differences can create a disconnect between the virtual objects and physical surfaces, which can break the illusion of realism for the viewer. Until smartphone devices with LIDAR sensing become commonplace and users are willing and able to share their entire physical environment, one way of mitigating the challenge of spatial dependencies is to guide the story consumer to arrange their physical environment to match what the creator had in mind more closely, something people usually do for watching football games together. SceneAR does this through the story description. 
\subsection{Include Spatial Navigation}
The story consumer has complete control over how and where they view the AR scenes. However, this can negatively impact the viewing experience owing to accidental camera movements. For example, if the person places a scene on the floor and happens to turn around, that scene will fall out of the camera's viewport. In such scenarios, locating the scene again with the camera's viewfinder, without virtual visual guidance, may be difficult. One way to resolve this is to provide directional arrows that point the viewer toward the center of the scene, especially if they are noticeably off-track from finding the scene with their camera. More complex 3D navigation information is needed to guide the viewer in multi-plane AR stories. 

\subsection{Recognize Camera Clutter} 
The ease with which creators can add virtual objects to a scene can lead them to crowd the camera's viewport and make it difficult to interact with the objects in the scene (e.g., selecting, scaling, moving). Some of our participants reported this problem. 
Designers can mitigate this problem by exploring different narrative structures, such as timed objects or plane-to-plane movement, to prevent AR objects from overcrowding a single viewport.

\subsection{Enable Support for Contextual Constraints} 
The differences between the creator's and viewer's environment introduce some semantic challenges in addition to the spatial ones described above. For example, a viewer would miss some of the story's context if they saw the story about the two penguins (Figure~\ref{fig:tub}) in a place without a bathtub. It would not present exactly as the creator intended. Again, once LIDAR sensors in mobile devices become commonplace, sharing the story context in 3D will become simpler but might introduce its own set of challenges like privacy. Meanwhile, a simple way to alleviate this challenge is by listing physical dependencies of the stories and leaving the consumer in charge of their AR experience of that story.
Effective storytelling in AR truly augments a user's reality; hence, describing the type of reality the creator intended to augment is necessary until we are able to easily create and share the 3D context with each AR narrative.

\subsection{Assess Surface Detection Limitations} Unlike traditional touchscreen apps, creating and viewing stories in AR requires physical movement to scan the physical space, detect planes, and move around for an immersive experience. Several factors can impact surface detection, including the user's device, lighting, clutter, and textures (or lack thereof). Repeated surface detection failures can make the AR experience feel onerous. One way to mitigate this issue, for the creator and the viewer, is to often save or cache the story content to prevent the story from disappearing when the detected surface is lost. While newer Android devices can support Depth APIs \cite{depthapi,arkit} and newer Apple devices have LIDAR sensors, both of which can improve the user's AR experience, most people do not own these devices yet, especially if we want to consider bringing AR to the billions of mobile devices in use worldwide. 

\subsection{Offer Help with Writer's Block} 
People can feel intimidated when creating a story from scratch, as some of our participants reported. Although this is not unique to AR, the spatial nature and even the novelty of AR might exacerbate this problem. One way to mitigate this is to scaffold new stories through previous stories, such as by allowing people to remix other people's stories as supported in SceneAR. Other options include offering in-app suggestions or templates. For example, the app can detect the user's location (e.g., a public park) and offer a story template based on it, or, as one participant suggested, the app can offer suggestions for virtual objects based on physical object recognition in their immediate environment. Although suggestions can be helpful, systems that scaffold (e.g., for novices) and support open-ended explorations for experts would allow us to build on decades of user-centered research in the design of web and mobile interfaces. Interestingly, participants commented that SceneAR could be the next AR meme generation app because of its rich data set of models and the freedom to create, share, and modify combinations of visuals and text.

\section{Limitations}
Our research provides insights for using AR as a storytelling platform for creating scene-based micro narratives. However, our work has some limitations. We focused on understanding how people would use an AR storytelling app and the opportunities and challenges that would arise. We chose to evaluate SceneAR as a minimal viable prototype, directing the user's efforts toward the creating, publishing, viewing, and remixing of micro stories rather than optimizing elements of the UI or adding more complex features. With informal testing, and later with the pilot study, we realized that users are not familiar with AR; namely, they do not understand how to scan, what types of surfaces make good candidates for plane detection (e.g., not blank walls or floors or white tables), nor do they realize the impact of low lighting levels on surface detection and tracking, to name a few. Therefore, while our original application had complex elements like hand gesture recognition and finger-based drawing in space, we chose to keep things simpler for the evaluation. We plan to investigate the multi-modal input modalities (gesture, speech, and touch) in a more controlled lab environment once we are allowed to do so in the fall. 
Although conducting the study with remotely located participants increased the complexity of the study design, we believe it also helped us learn things about AR use we may not have learned in an in-lab study, as mentioned above.

Additionally, the demographics of our sample may have biased our results. While the average age of our participants might align with those who are more likely to use creative or social media apps, our sample does not include older adults or children (i.e., those 18 years and below). A future study of two larger groups of users---with and without storytelling or creative backgrounds---may help us better understand narrative strategies and functionalities of our system beyond the end user social scenario considered in this work.

\section{Conclusion}
We presented SceneAR, a mobile AR app that enables users to create, publish, view, and remix scene-based micro narratives in AR. We detailed the design and implementation of SceneAR and presented findings from a three-day field study with 18 participants. Remixing enabled a new form of visual communication between participants, both by modifying a story and watching others modify and remix their stories. Conducting the study outside the lab environment revealed design tensions in AR apps primarily due to ARCore issues related to reliable surface detection and environmental conditions like lighting, surface textures, and clutter. Participants expressed a desire for a social network where they can create longer story-based communication threads, in both private and public conversations, or use the app to create AR memes. Based on the study, we outlined design strategies for AR scene-based narrative systems, highlighting characteristics unique to smartphone-based AR.

\bibliographystyle{abbrv-doi-hyperref}

\bibliography{template}
\end{document}